\documentclass[aps,prl,twocolumn]{revtex4-2} 

\usepackage[colorlinks,hyperindex,plainpages=false,allcolors={blue}]{hyperref}
\usepackage{longtable}
\usepackage{morefloats}
\usepackage{color}
\usepackage{amsmath}
\usepackage{graphicx,amsfonts,amsbsy}
\usepackage{amsmath,amsfonts,amsthm,amssymb}
\usepackage{appendix}
\usepackage{makeidx}
\usepackage{url}
\usepackage{verbatim}
\usepackage[rightcaption]{sidecap}
\usepackage{array}
\usepackage{booktabs}
\usepackage{multirow}
\usepackage{tabularx}
\usepackage{mathrsfs}
\usepackage{soul,xcolor}
\usepackage{times}
\usepackage{txfonts}
\usepackage{bm}

\newcommand{\be}{\begin{equation}}
\newcommand{\ee}{\end{equation}}
\newcommand{\ben}{\begin{eqnarray}}
\newcommand{\een}{\end{eqnarray}}

\begin{document}

\title{Magnon-Plasmon Hybridization Mediated by Spin-Orbit Interaction in Magnetic Materials}

\author{Anna Dyrda\l$^{1}$}
\email{adyrdal@amu.edu.pl}

\author{Alireza Qaiumzadeh$^{2}$}
\email{alireza.qaiumzadeh@ntnu.no}

\author{Arne Brataas$^{2}$}

\author{J\'ozef Barna\'s$^{1,3}$}

\affiliation{$^{1}$Faculty of Physics, Adam Mickiewicz University in Pozna\'n, ul. Uniwersytetu Pozna\'nskiego 2, 61-614 Pozna\'n, Poland\\
$^{2}$Center for Quantum Spintronics, Department of Physics, Norwegian University of Science and Technology, NO-7491 Trondheim, Norway\\
$^{3}$Institute of Molecular Physics, Polish Academy of Sciences, ul. M. Smoluchowskiego 17, 60-179 Pozna\'{n}, Poland}

\date{\today}
\begin{abstract}
We propose a mechanism for magnon-plasmon coupling and hybridization in ferromagnetic (FM) and antiferromagnetic (AFM) systems. The electric field associated with plasmon oscillations creates a non-equilibrium spin density via the inverse spin galvanic effect. This plasmon-induced spin density couples to magnons by an exchange interaction. The strength of magnon-plasmon coupling depends on the magneto-electric susceptibility of the system and the wavevector at which the level repulsion is happened. This wavevector may be tuned by an applied magnetic field. In AFM systems, the degeneracy of two chiral magnons is broken in the presence of a magnetic field, and we find two separate hybrid modes for left-handed and right-handed AFM magnons. Furthermore, we show that magnon-plasmon coupling in AFM systems is enhanced because of strong intra-sublattice spin dynamics. 
We argue that the recently discovered two-dimensional magnetic systems are ideal platforms to investigate proposed magnon-plasmon hybrid modes.
\end{abstract}
\maketitle

{\it{Introduction--}}
Collective excitations in condensed matter systems are emergent phenomena arising from many-body interactions. For example, three fundamental collective excitations in crystals are phonons (quanta of lattice oscillations), magnons (quanta of spin oscillations), and plasmons (quanta of charge-density oscillations) \cite{Atland,VignaleBook,Gross}. Interaction between these three bosonic excitations, at the lowest order of interaction, leads to the hybridization of two modes and, in higher orders, leads to various scattering phenomena. 
The Hybridization of two bosonic modes manifests as an energy-level repulsion, giving rise to an anti-crossing gap at the intersection in the dispersion curves.
Hybridization of the modes is interesting since it results in various topologically trivial and non-trivial emergent modes and may reveal some information about the quantum and topological nature of the system. 
Phonon-magnon \cite{magnon-phonon1,magnon-phonon2,magnon-phonon3,magnon-phonon4,magnon-phonon5,magnon-phonon6,magnon-phonon7} and phonon-plasmon \cite{plasmon-phonon1,plasmon-phonon2,plasmon-phonon3} hybrid modes are among the most studied hybrid modes in the previous decades. 

However, magnon-plasmon hybrid modes have received less attention, so far \cite{Baskaran,JB1,JB2,JB3}.
Hybridization of magnon-plasmon modes needs fine-tuning the matching frequency and wavevector of two modes and an effective interaction between them that leads to the anti-crossing gap at the intersection in the dispersion curves. In three-dimensional (3D) metallic systems, the plasmon mode has an intrinsic band gap of optical frequency \cite{VignaleBook}. At the same time, magnons operate at GHz and THz regimes in ferromagnetic (FM) and antiferromagnetic (AFM) systems, respectively \cite{RezendeJAP2019,akhiezer1968spin}. Therefore, it is hard to achieve the frequency-matching criteria. On the other hand, it is known that plasmon dispersion in two-dimensional (2D) systems is gapless \cite{VignaleBook}. The discovery of 2D graphene layers \cite{graphene} and surface states of topological insulators \cite{TI2,TI3,TI1} make it possible to investigate magnon-plasmon hybrid modes in heterostructure of 2D metal and magnetic insulator bilayers. In a recent theoretical study, a topologically non-trivial magnon-plasmon hybrid mode at the interface of a topological insulator--FM insulator bilayer was proposed, leading to a large thermal Hall response \cite{Kargarian}.

In this Letter we propose a new mechanism of magnon-plasmon hybridization based on  the electronic spin-orbit coupling (SOC) and {\it{s}}--{\it{d}}({\it{f}}) exchange interaction.
We argue that the plasmon oscillations may induce a non-equilibrium spin density via inverse spin galvanic effect or Edelstein effect \cite{Dyakonov,Ivchenko,Aronov,Edelstein,Ganichev,Kato04,Wang,Dyrdal,Raimondi}. Hence, the plasmon-induced spin density is coupled to magnon modes via {\it{s}}--{\it{d}}({\it{f}}) exchange interaction. In this scenario the magnon-plasmon coupling strength  is linearly proportional to the wavevector. 
We show that the proposed here mechanism  is very general and applies to 3D magnetic semiconductors and 2D metallic FM and AFM systems. 
As the recent discoveries of 2D magnetic systems opened a new path toward exploring emergent quantum many-body effects in low-dimensional magnetic systems \cite{2Dmagnets,2Dmagnets1}, we focus here mainly on 2D FM and AFM systems. In particular the 2D metallic AFM systems, that support two chiral magnon modes with opposite spin polarizations, seem to be exciting candidates for exploring novel magnon-plasmon hybrid modes.

{\it{General formalism of magnon-plasmon hybridization--}}
First, we should obtain an effective magnon-plasmon Hamiltonian based on our proposal. To do that, we consider a metallic low-symmetry magnetic system with the following interacting Hamiltonian,
\begin{align}\label{Hint}
H=H_{\rm el} + H_{\rm S} + H_{\rm int}.
\end{align}
The  first term, $H_{\rm el}$, describes the electronic subsystem and includes kinetic term, Coulomb interaction, and SOCs. The second term, $H_{\rm s}$, represents the FM or AFM magnetic subsystem and generally includes Heisenberg exchange interactions, magnetic anisotropies, dipolar interactions, and Dzyaloshinskii–Moriya interactions (DMIs). Finally, the last term in the Hamiltonian describes the coupling between these two subsystems that is modeled by a Zener-type {\it{s}}--{\it{d}}({\it{f}}) exchange interaction between the spin of the conducting {\it{s}}-orbital electrons ${{\bf s}_i}$ and localized {\it{d}}- ({\it{f}}-) orbital electrons ${\bf S}_i$ at site $i$, 
\begin{align} \label{sd}
    H_{\rm int}= -I_{0} \sum_i {\bf S}_i \cdot  {{\bf s}_i}
\end{align}
where $I_{0}$ parametrizes the strength of interaction \cite{Zener,Yosida_Magnetism,vonsovsky1974magnetism}. 

We are interested in the plasmon contribution of the electronic Hamiltonian, $H_{\rm el}$. By introducing collective coordinates for the long-range part of the Coulomb interactions \cite{Pines, Gross}, the Hamiltonian of interacting electrons can be transformed into an effective Hamiltonian that consists of terms describing a short-range interacting electron liquid, free bosonic plasmons, and electron-plasmon interactions \cite{Pines, Gross}. The electron-plasmon term is important as it leads to strong so-called Landau damping of plasmons by creating electron-hole pairs when plasmon dispersion enters the electron-hole continuum of the electron liquid at a critical wavevector ${\bf q}_c$. Here, we only consider free and undamped plasmons, and thus the interacting electronic Hamiltonian $H_{\rm el}$, reduces to the following plasmon Hamiltonian \cite{Pines,Gross,Overhauser},
\begin{equation}
\mathcal{H}_{\rm pl}=\sum_{{{\bf q} < {\bf q}_c}}\hbar \omega_{\rm pl}a^{\dagger}_{\bf q}a_{\bf q} \, , 
\end{equation}
where $a^{\dagger}_{\bf q}$ ($a_{\bf q}$) is the bosonic creation (annihilation) operator of a plasmon mode with a wavevector ${\bf q}$, $\omega_{\bf q}$ is the corresponding plasmon frequency, and $\hbar$ is the reduced Planck constant. Within the Random Phase Approximation (RPA), the plasmon dispersion at long wavelengths in 3D electron liquids is given by  $\omega_{\rm pl}\simeq\Omega_0\big(1 + 3 v^2_F q^2/10 \Omega_0^2 \big)$, while in 2D systems, we have $\omega_{\rm pl} \simeq \Omega_0 \sqrt{q/2}$ \cite{VignaleBook,Polini0,Polini,Maslov}. Here $\Omega_0=\sqrt{4\pi n e^2/m}$, $v_F$ is the Fermi velocity, $n$ is the charge carrier density, $e$ is the electron charge, and $m$ is the electron effective mass. 
In 2D systems, the plasmon dispersion is gapless $\omega_{\rm pl}({\bm q} \rightarrow 0)=0$, while 3D systems have an intrinsic plasmon gap of $\Omega_0$ that depends on the charge density. 

In the magnetic subsystem $H_{\rm S}$, we are interested in the low-energy spin excitations, called magnons. Using the Holstein-Primakoff bosonization technique and within the linear spin-wave theory, the spin Hamiltonian $H_{\rm S}$ reduces to the following free magnon Hamiltonian for FM and AFM systems \cite{akhiezer1968spin},
\begin{align}
 \mathcal{H}^{\rm FM}_{\rm m} &= \sum_{\bf q} \hbar \omega_{\rm m} b^{\dagger}_{\bf q} b_{\bf q},\\
 \mathcal{H}^{\rm AFM}_{\rm m} &= \sum_{\bf q , \sigma} \hbar \omega_{\rm m}^{\sigma} b^{\dagger}_{\bf q \sigma} b_{\bf q \sigma}.
\end{align}
Here $b^{\dagger}_{\bf q}$ ($b_{\bf q}$) and $b^{\dagger}_{\bf q \sigma}$ ($b_{\bf q \sigma}$) are, respectively, boson creation (annihilation) operators at wavevector $\bm q$ in FM and AFM systems, with corresponding dispersion $\omega_m$ and $\omega_{m}^{\sigma}$, respectively. AFM systems commonly have two magnon eigenmodes with right-/left-handed spin polarization (chirality), denoted by $\sigma=\uparrow$/$\downarrow$,  while magnons in FM systems are only right-handed. The degeneracy of two AFM magnon modes can be broken by applying a magnetic field.

Now, we formulate the effective Hamiltonian of  magnon-plasmon coupling. 
As plasmons are  associated with space-time oscillations of the charge density, they inherently generate an oscillating longitudinal electric field. This electric field can couple to the magnetic subsystem {\it via } SOCs, which effectively leads to the magnon-plasmon interaction. In fact, 
coupling of plasmons to magnons can be mediated either through the SOC in the electronic subsystem or through the SOC in the spin subsystem. In the former case, the electric field due to plasmons generates a dynamical spin polarization of the charge carriers via inverse galvanomagnetic effect, and resulting non-equilibrium spin polarization can be coupled to the localized spins through the {\it{s}}--{\it{d}}({\it{f}}) exchange interaction $H_{\rm int}$. In the second case, the plasmon electric field leads to dynamical spin polarization of the spin subsystem via SOC, which effectively gives rise to magnon-plasmon coupling. 
In the following, we focus on the magnon-plasmon coupling due to SOC in the electronic subsystem.

The longitudinal electric field associated with plasmon oscillations in $d=\{2, 3\}$ dimensions can be computed using a method introduced in Ref.~\cite{Deigen},  
\begin{align}\label{electric}
{\bm E} =\frac{2^{d-1}\pi ne}{\epsilon} \left(\frac{\hbar}{2 L^d n m}\right)^{1/2} \sum_{\bf q}\frac{{\bf q}}{q^{d-2} \omega^{1/2}_{\rm pl}} (a^{\dagger}_{-\bf q}-a_{\bf q})e^{i{\bf {q}}\cdot {\bf {r}}},
\end{align}
where $L$ is the system size and $\epsilon$ is the material dielectric constant.  

In the linear response regime, an ac electric field of frequency $\omega$ induces a non-equilibrium ac spin polarization via the inverse spin galvanic effect,
\begin{align} \label{susceptibility}
\delta s^a_{\omega} &= \sum_j\chi^{ab}_{\omega} E^b_{\omega},
\end{align}
where $\chi^{ab}_{\omega}$ is the dynamical magneto-electric susceptibility or spin-charge response function of the electronic subsystem, with $a,b=\{x,y,z\}$. 
This response function may have an extrinsic contribution, proportional to the electron's relaxation time, and/or intrinsic contribution, arising from the Berry curvature of electronic bands \cite{Alireza3}.
Therefore, the SOC acting in the conducting electron subsystem convert  the plasmon-induced electric field, Eq. (\ref{electric}), to a non-equilibrium spin density, Eq. (\ref{susceptibility}). This induced ac spin density interacts with magnon excitations via {\it s}--{\it d}({\it f}) interaction, Eq. (\ref{sd}). Therefore, we can finally obtain the lowest order effective Hamiltonian of the magnon-plasmon interaction in FM and AFM metals as, 
\begin{align}
\label{FM-pl}
&\mathcal{H}^{\rm FM}_{\rm m-pl}  =\sum_{\bf q} \hbar (a_{\bf q}-a^+_{-\bf q})
\left[ \mathcal{C}^{\rm FM}_{\bf q}b_{-\bf q}
+ \mathcal{C}^{*\rm FM}_{\bf q}b^{\dagger}_{\bf q} \right],\\
\label{AFM-pl}
&\mathcal{H}^{\rm AFM}_{\rm m-pl}  =\sum_{\bf q} \hbar (a_{\bf q}-a^{\dagger}_{-\bf q})
\left[ \mathcal{C}^{\rm AFM}_{\bf q}(b_{ -{\bf q}\downarrow}+b^{\dagger}_{{\bf q}\uparrow })+ \mathcal{C}^{*\rm AFM}_{\bf q}(b^{\dagger}_{ {\bf q}\downarrow}+b_{-{\bf q}\uparrow }) \right],
\end{align}
where $\mathcal{C}^{\rm FM}_{\bf q}=\mathcal{B}_{\bf q} \mathcal{F}_{\bf q}/\hbar$ and $\mathcal{C}^{\rm AFM}_{\bf q}=(u_{\bf q}+v_{\bf q})\mathcal{B}_{\bf q} \mathcal{F}_{\bf q}/\hbar$ are the effective magnon-plasmon coupling strength of FM and AFM systems, respectively. We define ${\mathcal{F}_{\bf q}=F_{\bf q}^x-iF_{\bf q}^y}$, where ${\mathcal{F}^a_{\bf q} = \sum_b q_{b} \chi^{ab}_{\omega=\omega_{\rm  pl}}}$, and ${\mathcal{B}_{\bf q}}$ in 3D is given by
${ \mathcal{B}^{\rm 3D}_{\bf q} = I_0\sqrt{\pi\hbar Sn_s\omega_{\rm pl}}(\epsilon q)^{-1}}$ while in 2D is
$ {\mathcal{B}^{\rm 2D}_{\bf q} =
 {I_0}{\epsilon}^{-1}\sqrt{{\pi \hbar Sn_s\omega_{\rm pl}}/{2q}}}$, with $S=|\bm{S}_i|$ denoting the spin length and $n_s$ the number of the lattice sites per unit cell.  The magnon-plasmon coupling constant in AFM systems is enhanced by a factor of $(u_{\bf q}+v_{\bf q})$, where $u_{\bf q}$ and $v_{\bf q}$ are AFM Bogoliubov transformation coefficients \cite{RezendeJAP2019,Alireza2021}. This enhancement is attributed to the strong exchange-dominant intra-sublattice dynamics of two AFM spins in a magnetic unit cell.

Eventually, the total Hamiltonian of the system, Eq. (\ref{Hint}), is reduced to the following effective Hamiltonian of interacting FM (AFM) magnon and plasmon collective modes,
\begin{align}\label{Heff}
\mathcal{H}^{\rm FM(AFM)} = \mathcal{H}_{\rm pl} + \mathcal{H}^{\rm FM(AFM)}_{\rm m} + \mathcal{H}^{\rm FM(AFM)}_{\rm m-pl}.
\end{align}

{\it{Magnon-plasmon hybrid modes in generic 2D systems--}}
The proposed magnon-plasmon coupling mechanism, Eq. (\ref{Heff}), is quite generic and can be applied in 3D and 2D magnetic systems. However, as we mentioned earlier, the plasmon dispersion in 3D systems has an intrinsic gap of optical frequency and can hardly be hybridized with FM and AFM magnons that are in GHz and THz regime, respectively. 
On the other hand, plasmons in 2D systems are soft modes with a tunable energy dispersion \cite{2Dmagnets1}, hence, recently discovered 2D magnetic materials are promising candidates for exploring magnon-plasmon hybrid modes.  Therefore, in the rest of this Letter, without loss of generality, we assume 2D FM and AFM metallic systems with square lattice structure.  Accordingly, the spin Hamiltonian of the magnetic subsystem is
\begin{equation}
\mathcal{H}_{\rm S} = \mp J \sum_{\langle i j \rangle} \mathbf{S}_i \cdot \mathbf{S}_j - K_z \sum_{i} ( S^z_i)^2+ g \mu_B  H_0 \sum_{i}S^z_i,
\end{equation}
where $\langle ij \rangle$ denotes summation over nearest-neighbor sites, $i$ and $j$, $J >0$ represents the isotropic exchange interaction, and the sign $\mp$ in front of $J$ corresponds to  FM/AFM ordering, respectively. Furthermore, ${K}_z  > 0$ is the anisotropy constant, $H_0$ is the magnetic field along the $z$ direction, $g$ is the Land\'e factor, and $\mu_B$ is the Bohr magneton. The dispersion of FM and AFM magnons read,
\begin{align}
\label{FMmagnon}
\hbar \omega_{\rm m} &=g\mu_B (H_0+H_A) +zJ S(1-\gamma_{\bf{k}}),\\
\label{AFMmagnon}
\hbar \omega_{\rm m}^{\uparrow,\downarrow} &= \sqrt{ (zJS+H_A)^2-(zJS\gamma_{\bm q}) ^2} \mp g \mu_B H_0,
\end{align}
where $\gamma_{\bm q}=z^{-1} \sum_{\bm \delta} \exp{(i \bm{q}\cdot\bm{\delta})}$ is the lattice structure factor, with $z$ and $\bm \delta$ denoting the coordination number ($z=4$ for 2D square lattice) and nearest-neighbor vectors, respectively, while $H_A=2K_zS/g\mu_B$ is the anisotropy field.   

To compute the magnon-plasmon hybrid modes, we should first find the effective magnon-plasmon coupling strength, $\mathcal{C}_{\bf q}$ in Eqs. (\ref{FM-pl}) and (\ref{AFM-pl}) that is proportional to $\mathcal{B}_{\bf q}\mathcal{F}^a_{\bf q}$ in both FM and AFM systems.
$\mathcal{B}_{\bf q}$ is linearly proportional to the {{\it s}--{\it d}({\it f})} exchange interaction $I_0$ and $\mathcal{F}^a_{\bf q}$
is related to the dynamical magneto-electric susceptibility, $\chi^{ab}_{\omega}$, of the magnetic system. In the following, for numerical calculations, we compute the effective magnon-plasmon coupling up to the linear order in $I_0$. Therefore, the magneto-electric susceptibility can be calculated in the nonmagnetic limit $\chi^{ab}_{\omega}(I_0 \rightarrow 0)$. 
On the other hand, the induced spin polarization in 2D nonmagnetic materials is  perpendicular to the applied electric field direction and is proportional to the electron relaxation time (see, e.g.,\cite{Edelstein,Dyrdal,Raimondi,Alireza3}). Therefore, in the frequency region of the interest, $\omega_{\rm pl} \tau \ll 1$, where $\tau$ is the electron scattering time, it is approximately frequency independent.

{\it{2D FM magnon-plasmon hybridization--}}
First, we find the dispersion relation of  magnon-plasmon modes in FM case.
The total FM magnon-plasmon Hamiltonian, Eq. (\ref{Heff}), can be written as ${\mathcal{H}}^{\rm FM}=\sum_{\bf{q}}\Phi^{\dagger}_{\bf{q}}\mathbb{H}_{\bf q}^{\rm FM}\Phi_{\bf{q}}$, with the vector field operator $\Phi_{\bf{q}}=(a_{\bf{q}},b_{\bf{q}},  a^{\dagger}_{-\bf{q}},b^{\dagger}_{-\bf{q}})^T$, and $\mathbb{H}_{\bf q}^{\rm FM}$ defined as,
\begin{eqnarray}
\mathbb{H}_{\bm q}^{\rm FM} =
\hbar\left(
   \begin{array}{cccc}
  \omega_{\rm pl} & \mathcal{C}_{\rm FM} & 0 & \mathcal{C}^{*}_{\rm FM} \\
   \mathcal{C}^{*}_{\rm FM} & \omega_{\rm m} &  -\mathcal{C}^{*}_{\rm FM} & 0 \\
   0 & - \mathcal{C}_{\rm FM} & \omega_{\rm pl} &  - \mathcal{C}_{\rm FM} \\
   \mathcal{C}_{\rm FM} & 0 &  -\mathcal{C}^{*}_{\rm FM} & \omega_{\rm m} \\
\end{array}
\right).
\end{eqnarray}
For clarity reasons, the q-dependence  of the coupling parameter has been suppressed here.
This bosonic  Hamiltonian  is now diagonalized using the  procedure described in Ref.
[\onlinecite{White,RezendeJAP2019}], and we find the following dispersion relations for FM magnon-plasmon hybrid modes, 
\begin{align} \label{FM-plasmon}
\omega^{1,2}_{\rm{m-pl}}=\frac{1}{\sqrt{2}}\sqrt{\omega_{\rm pl}^2+\omega_{\rm m}^2\pm \sqrt{(\omega_{\rm pl}^2-\omega_{\rm m}^2)^2+16 |\mathcal{C}_{\rm FM}|^2 \omega_{\rm pl} \omega_{\rm m}}}
\end{align}
In the absence of magnon-plasmon coupling, $\mathcal{C}_{\rm FM}=0$, the above relations   reduce  to those of decoupled   magnon and plasmon modes.

\begin{figure*}
\includegraphics[width=2.0\columnwidth]{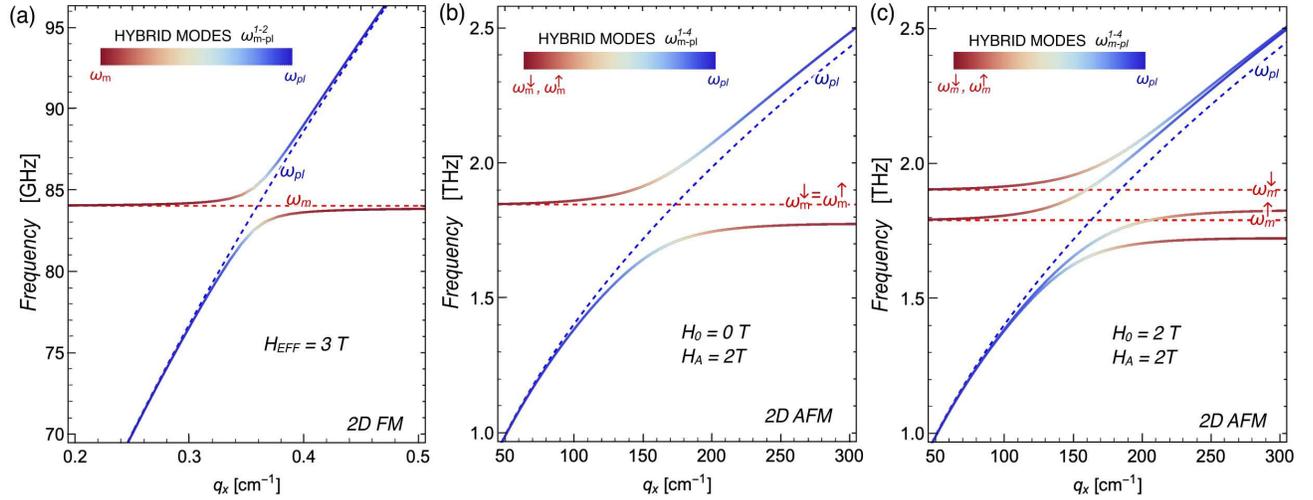}
\caption{Magnon-plasmon hybridization in a 2D FM system (a), and a 2D AFM system in the absence (b) and presence (c) of a magnetic field. Red and blue dashed lines present decoupled magnon and plasmon eigenmodes, respectively. Solid lines are hybridized magnon-plasmon modes. $H_{\rm eff}$ in (a) is a sum of external and anisotropy fields, $H_{\rm eff} =H_0+H_A$. The other parameters are: $J = 5$meV, $I_{0} = 3.6\cdot 10^{-15} \,{\rm meVcm^{2}}$, $m=0.9 m_0$,$n = 1.1 \cdot 10^{13}\, {\rm cm^{-2}}$, $n_s = 1.1 \cdot 10^{15}\, {\rm cm^{-2}}$, $\epsilon = 1$, while $\chi_{xy} = 10^{12} \, {\rm s/\sqrt{{\rm cm}^{3}{\rm g}}}$. }
 \label{fig:Figall}
\end{figure*}

Figure \ref{fig:Figall}(a) shows that the dispersion curves of non-interacting magnons and plasmons in a 2D FM system have an intersection at certain wavevector and frequency.
Upon turning the magnon-plasmon coupling on, $\mathcal{C}_{\rm FM} \neq 0$, the hybrid magnon-plasmon states are formed around the intersection, that manifests as a level repulsion (level anticrossing) of the two modes.
The  magnon gap in FM system can be tuned by an external magnetic field, Eq. (\ref{FMmagnon}). Thus, the frequency and magnitude of the anti-crossing gap can be tuned, as well. 

{\it{2D AFM magnon-plasmon hybridization--}}
2D AFM systems are more interesting since there are two polarized magnon modes, and by an applied magnetic field, one can tune the hybridization of two magnon modes with plasmons.
The effective AFM Hamiltonian, Eq. (\ref{Heff}), can be written as ${\mathcal{H}}^{\rm AFM}=\sum_{\bf{q}}\Psi^{\dagger}_{\bf{q}}\mathbb{H}_{\bf q}^{\rm AFM}\Psi_{\bf{q}}$, with the vector field operator $\Psi_{\bf{q}}=(a_{\bf{q}},b_{\bf{q}\uparrow},  a^{\dagger}_{-\bf{q}},b^{\dagger}_{-\bf{q}\downarrow})^T$, and $\mathbb{H}_{\bf q}^{\rm AFM}$ given by,
\begin{eqnarray}
\mathbb{H}_{\bm q}^{\rm AFM} =\hbar
\left(
   \begin{array}{cccc}
   \omega_{\rm pl} & \mathcal{C}^{*}_{\rm AFM} & 0 & \mathcal{C}^{*}_{\rm AFM} \\
   \mathcal{C}_{\rm AFM} & \omega_{\rm m}^{\uparrow} &  -\mathcal{C}_{\rm AFM} & 0 \\
   0 & - \mathcal{C}^{*}_{\rm AFM} & \omega_{\rm pl} &  - \mathcal{C}^{*}_{\rm AFM} \\
   \mathcal{C}_{\rm AFM} & 0 &  -\mathcal{C}_{\rm AFM} & \omega_{\rm m}^{\downarrow} \\
\end{array}
\right).
\end{eqnarray}
We should solve the following quartic equations for the hybrid modes, $\omega = \omega_{\rm m-pl}^{1-4}$:
\begin{align} 
(\omega^2-\omega_{\rm pl}^2)& (\omega \pm \omega_{\rm m}^\downarrow)(\omega\mp\omega_{\rm m}^\uparrow) 
- 2 |\mathcal{C}_{\rm AFM}|^2 \omega_{\rm pl} (\omega_{\rm m}^\downarrow+\omega_{\rm m}^\uparrow)=0.
\end{align}
The general solutions of these two equations for non-degenerate AFM magnons are lengthy, and we do not represent them here. However, if AFM magnon modes are degenerate,  $\omega_{\rm m }^\downarrow=\omega_{\rm m }^\uparrow$
, the form of AFM magnon-plasmon dispersion is similar to the FM case, see Eq. (\ref{FM-plasmon}).

Figure \ref{fig:Figall}(b) and \ref{fig:Figall}(c) represent the dispersion curves of the non-interacting magnons and plasmons in a 2D AFM system. The degeneracy of AFM magnon modes is broken in the presence of external magnetic field, Eq. (\ref{AFMmagnon}), and thus the plasmon curve can intersect AFM magnon bands in two separate points with different frequencies and wavevectors, see Fig. \ref{fig:Figall}(c). Again upon turning the magnon-plasmon on, $\mathcal{C}_{\rm AFM}\neq 0$, the corresponding hybrid magnon-plasmon states are appeared as anti-crossing level repulsion of bands around the intersection curves.
Therefore, we can have two separate magnon-plasmon hybrid modes with opposite chirality at different frequencies and wavevectors, see Fig. \ref{fig:Figall}(c). Furthermore, since the strength of the magnon-plasmon coupling is proportional to the wavevector hence, the right-handed magnon mode $\omega^{\downarrow}_{\rm AFM}$ has a stronger interaction with plasmon mode and hence a larger anti-crossing gap than the left-handed magnons $\omega^{\uparrow}_{\rm AFM}$.    

{\it{Summary--}}
We have formulated a magnon-plasmon hybridization mechanism in FM and AFM systems. The hybridization in this model is mediated by  SOC in the electronic subsystem. An electric field associated with plasmon oscillations induces a non-equilibrium spin polarization via the Edelstein effect or inverse spin galvanic effect. This plasmon-induced spin polarization may interact with the magnons via a Zener-like {\it s}--{\it d}({\it f}) coupling interaction. The strength of magnon-plasmon hybridization depends on the magneto-electric  susceptibility and {\it s}--{\it d}({\it f}) exchange interaction. 
We propose the recently discovered 2D FM and AFM systems are ideal candidates for observation of this effect. Also the interface of a magnetic insulator and a heavy metal may host magnon-plasmon hybrid mode associated with our proposed mechanism. In AFM systems, we can tune the band splitting of two chiral AFM magnon eigenmodes and thus adjust the frequency, wavevector, coupling strength, and polarization of the magnon-plasmon hybrid mode. We found an enhancement of magnon-plasmon coupling in AFM systems compared to their FM counterpart. This enhancement appears because two AFM sublattices are strongly entangled and involved in the magnon dynamics in AFM systems, and is described by a factor proportional to the corresponding Bogoliubov transformation coefficients. We believe that the magnon-plasmon hybridization will become an important issue in the following, as a connection of already well-developed plasmonics and magnonics.   \\

{\it{Note added--}}— Recently, we became aware of two papers 
that discussed magnon-plasmon hybridization in 2D FM systems. In Ref. \cite{Peres}, the mechanism of FM magnon-plasmon hybridization is based on the direct Zeeman coupling of the electromagnetic field of plasmon oscillations to the localized spins. However, in Ref. \cite{Katsnelson}, the hybridization mechanism is based on the spin polarization of the bands in 2D FM metals. 

{\it{Acknowledgments--}}- This work has been supported by the Norwegian Financial Mechanism under the Polish-Norwegian Research Project NCN GRIEG, project No. 2019/34/H/ST3/00515, '2Dtronics' (AD, AQ, JB);  AQ and AB acknowledge support of the Research Council of Norway
through the Centres of Excellence funding scheme, Project No. 262633, 'QuSpin'.


%

\end{document}